\documentclass[conference]{IEEEtran}

\usepackage{cite}
\usepackage{amsmath,amssymb,amsfonts}
\usepackage{algorithmic}
\usepackage{graphicx}
\usepackage{textcomp}
\usepackage{xcolor}
\usepackage{color,soul}
\usepackage[section]{placeins}
\usepackage{booktabs}
\usepackage{tabularx}
\usepackage[hidelinks]{hyperref}
\usepackage[capitalise]{cleveref}
\usepackage{inconsolata} 
\usepackage{fontawesome5}
\def\BibTeX{{\rm B\kern-.05em{\sc i\kern-.025em b}\kern-.08em
    T\kern-.1667em\lower.7ex\hbox{E}\kern-.125emX}}
\begin{document}

\title{Analyzing Logs of Large-Scale Software Systems using \textit{Time Curves} Visualization
}


\author{
	\IEEEauthorblockN{
		Dmytro Borysenkov\IEEEauthorrefmark{2},
		Adriano Vogel\IEEEauthorrefmark{2},
		S\"oren Henning\IEEEauthorrefmark{2}\IEEEauthorrefmark{3},
		Esteban Perez-Wohlfeil\IEEEauthorrefmark{2}\textsuperscript{\faIcon[regular]{envelope}}
		}
	\IEEEauthorblockA{
		\IEEEauthorrefmark{2}Dynatrace Research, Linz, Austria 
		\{dmytro.borysenkov, adriano.vogel, esteban.wohlfeil\}@dynatrace.com}
	\IEEEauthorblockA{
		\IEEEauthorrefmark{3}JKU/Dynatrace Co-Innovation Lab, Johannes Kepler University Linz, Austria
		\{soeren.henning\}@jku.at
	}
}

\maketitle

\begin{abstract}
Logs are crucial for analyzing large-scale software systems, offering insights into system health, performance, security threats, potential bugs, etc. However, their chaotic nature—characterized by sheer volume, lack of standards, and variability—makes manual analysis complex. The use of clustering algorithms can assist by grouping logs into a smaller set of templates, but lose the temporal and relational context in doing so. On the contrary, Large Language Models (LLMs) can provide meaningful explanations but struggle with processing large collections efficiently. Moreover, representation techniques for both approaches are typically limited to either plain text or traditional charting, especially when dealing with large-scale systems. In this paper, we combine clustering and LLM summarization with event detection and Multidimensional Scaling through the use of \textit{Time Curves} to produce a holistic pipeline that enables efficient and automatic summarization of vast collections of software system logs. The core of our approach is the proposal of a semimetric distance that effectively measures similarity between events, thus enabling a meaningful representation. We show that our method, based on logs collected from different applications, can explain the behavior of a system over time without prior knowledge. We also show how the approach can be used to detect general trends as well as outliers in parallel and distributed systems by overlapping multiple projections. As a result, we expect a significant reduction in the time required to analyze system-wide issues, identify performance bottlenecks and security risks, debug applications,\textit{ etc.}
\end{abstract}


\section{Introduction}
\label{Section_Introduction}

Analyzing log data presents significant challenges due to its poor formatting, immense volume and rapid growth \cite{ismail2017scaling}. At Dynatrace, we observe how software systems continue to expand, both the quantity of system logs and the rate at which they are generated are expected to increase \cite{dynatrace1MPerSecond}. Despite these challenges, log messages contain valuable information, such as error stack traces and execution details, that is often not documented elsewhere. Unfortunately, the semi-structured nature of system logs and their overwhelming volume make it difficult for humans and large language models (LLMs) to interpret and explain the monitored processes effectively \cite{liu2024lost}.

\begin{figure}[h]
\centering
   \includegraphics[scale=1.]{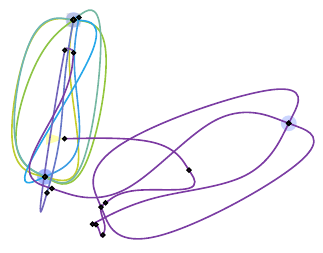}
\caption{Example \textit{Time Curve} corresponding to the analysis of the execution of a stream processing application, where data points represent events of the system, consecutive events in time are connected, while similar events are positioned close. Color of connection shows time, \textit{i.e.} start of observation purple -- end of observation green.}
\label{Figure_ExampleTimeCurve}
\end{figure}

Several sophisticated and automated solutions exist for tasks such as anomaly detection \cite{zhang2019robust} or failure prediction \cite{zheng2009system}, which have proven that log data can be useful. However, in practice, users typically examine system logs only after an automated alert is triggered, either because the issue cannot be resolved automatically or because they seek a deeper understanding of the problem. In both scenarios, the lack of explainability complicates log analysis. At Dynatrace, the immense volume of log ingestion and the need for real-time analytics have motivated us to explore visualization techniques that can summarize and explain the evolution of a software system using only log data. We believe that the suggested visualization can serve as an effective entry point for analyzing software systems by providing a comprehensive overview of system behavior and facilitating quick insights, thus enabling efficient exploration. Initially, one can quickly identify visual patterns and easily review underlying logs. If further investigation is needed, one can then dive into standard dashboards, integrating the insights gained (\textit{e.g.}, relevant time frame) with other available information (\textit{e.g.}, metrics). 

\cref{Figure_ExampleTimeCurve} illustrates the \textit{Time Curve} of the automatic analysis of the execution of a stream processing application \cite{fragkoulis2024survey} using the proposed method.

\begin{figure*}[h]
    \centering
    \includegraphics[width=0.95\textwidth]{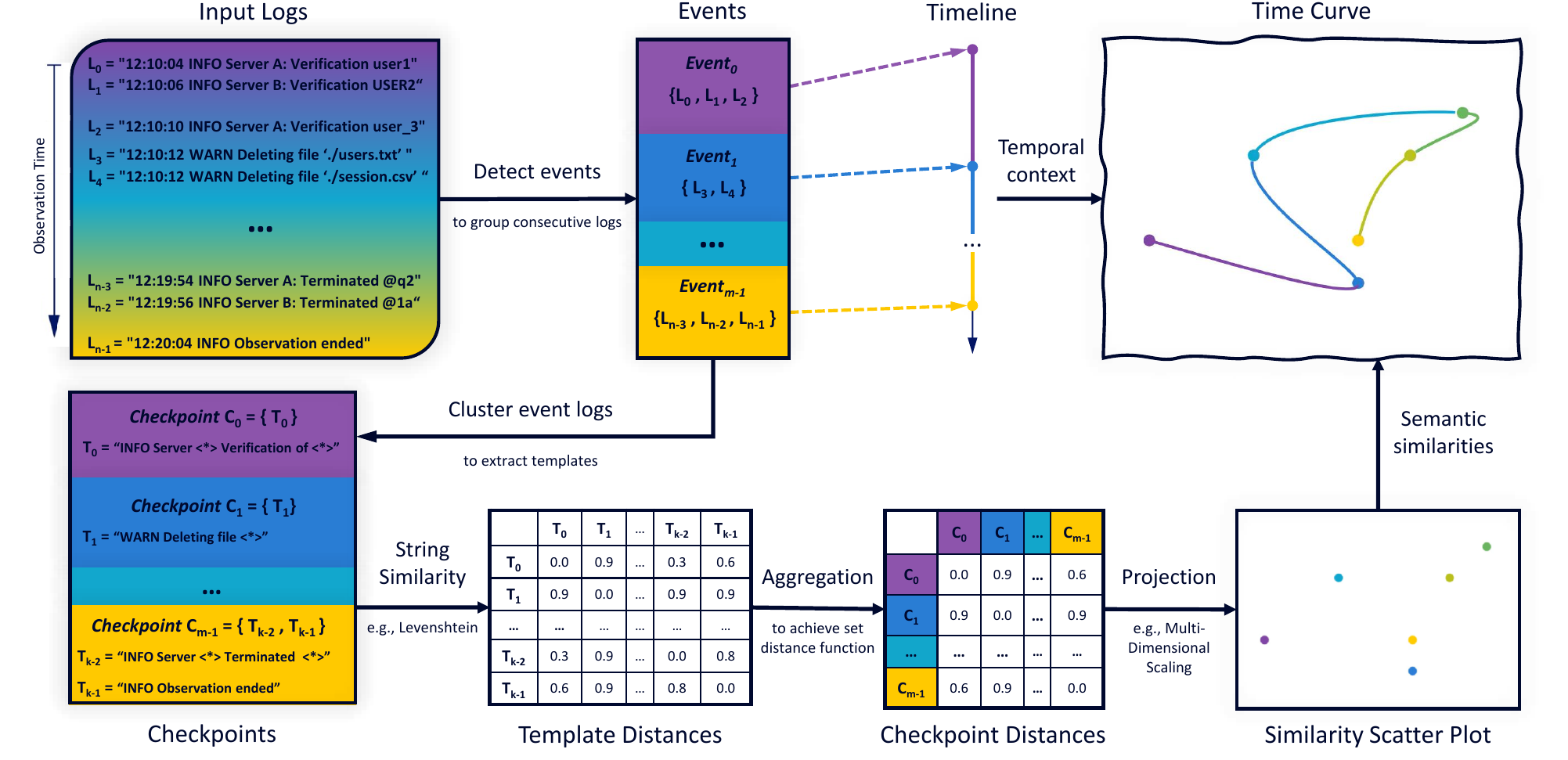}
\caption{Workflow illustration of the proposed method to generate a \textit{Time Curve} from a set of logs.}
\label{Figure_Pipeline}
\end{figure*}

Each data point corresponds to a system event, and the proximity between points depicts similarity between the corresponding sets of logs of the events. Although specific details on the analysis are given in \cref{Section_StreamProcessingAnalysis}, some visual patterns from \cite{bach2015timecurves} can already be intuitively perceived from the visualization itself, such as clusters of events (shown as halos indicating presence of almost identical events) and alternations (representing the repetitive nature of the system behavior). Note that a glossary containing definitions of visual patterns and other terms used in the paper is available in \cite{supplementaryMaterial}.

We propose a holistic pipeline (see \cref{Figure_Pipeline}) that enables to automatically extract processing trends and visualize the evolution of a system based exclusively on the underlying logs. Therefore, our approach can result in a significant reduction of the time required to analyze system outages, identify performance bottlenecks, perform root cause analysis, \textit{etc}., thus benefiting a wide spectrum of the industry. \\

Our proposal contributes several novel elements into the analysis tool set for system logs. From the technical perspective, we provide the following:
\begin{enumerate}
    \item An automatic event detection mechanism which groups large collections of system logs into a handful of events.
    \item A template extraction step via clustering to reduce dimensionality and increase scalability.
    \item A sensitive yet robust semimetric function to compute the similarity between sets of clustered logs.
    \item Several extensions to the \textit{Time Curves} \cite{bach2015timecurves} technique, including the comparison of multiple systems at once, animating curves or enriching them via LLM summarization.
\end{enumerate}

As a result of the experimental evaluation, the following scientific contributions are provided:

\begin{enumerate}
    \item The experimental approach as a validation and reference method for the analysis of log datasets. This is achieved by (1) using an experiment under controlled conditions, (2) a public dataset and (3) a performance benchmark.
    \item The experimental analysis and insights gained from the log datasets, which can be compared with further experiments and results from other tools.
\end{enumerate}

The remainder of this paper is structured as follows: \cref{Section_RelatedWork} discusses the current state-of-the-art of log analysis and visualization; \cref{Section_Methodology} introduces the concept of \textit{Time Curves} and the components of the processing pipeline (\cref{Figure_Pipeline}), such as event detection, clustering and distance computation; \cref{Section_ResultsAndDiscussion} describes the experiments performed to depict the capabilities of the method. Lastly, \cref{Section_Conclusions} concludes this study and discusses future work.

\section{Related work}
\label{Section_RelatedWork}

Dashboards are the most common tools for visualizing log data, typically featuring numerous detailed histograms and line plots that allow for interactive navigation, such as selecting time frames or applying filters. Although dashboards are theoretically well-suited for problem analysis, mastering them can be challenging, and advanced interactions require significant time investment. The extensive functionality of dashboards can overwhelm users and even cause confusion, especially among non-experts. However, an overall analysis of the evolution of the system can also be valuable for individuals with limited knowledge of the system, such as owners of small online services, salespeople, client support staff and developers from related domains. These points highlight the need for simple and intuitive visualization strategies.

For instance, Streamgraphs \cite{byron2008streamgraphs} or stacked bar charts can show composition over time (counts of similar logs over time), but require categorization of the data into distinct disjoint classes with certain properties of interest. In the case of logs this is not trivial since the most valuable information typically lies in the semantic meaning of log messages, which is hardly measurable.

Such lack of intuitive features can potentially be addressed through the use of embeddings. Visualizing the embedding space can provide insights into the composition of semantic content by training or utilizing pre-trained text embeddings and visualizing them, such as through down-projections on scatter plots \cite{huang2023embeddings}. While a robust embedding transformation can effectively display the distribution of log messages and their semantic similarities, it struggles to illustrate the evolution of the system due to the loss of temporal context. To incorporate chronological order, one can adapt trajectory-based methods similar to those in \cite{streit2020embeddingstrajectories}. However, in their approach, encodings need to be manually designed, which can be a challenging task depending on the data type.

Even with effective trajectory bundling as suggested in \cite{streit2020embeddingstrajectories}, the results can become visually complex. In practice, system logs can easily reach millions of lines and gigabytes of data. Therefore, grouping consecutive logs into events is crucial not only for visualization purposes but also due to computational limitations. For example, feeding an entire log file directly into any large language model (LLM) can be either unfeasible or overly expensive \cite{liu2024lost}. While utilizing the context window may suffice for tasks such as anomaly detection \cite{karlsen2024benchmarking}, it may be unsuitable for providing a comprehensive overview of the system over time. To enable AI assistance in analyzing the evolution of the system, we need to develop a more abstract and concise representation of the logs that still preserves all essential semantic information contained of the raw data.

Thinking in terms of events (groups of consecutive logs) allows for the consideration of visually simpler techniques, such as timelines. Timelines are easy to interpret and effectively display the flow of events over time. However, they do not incorporate semantic similarity, making the analysis challenging. One might attempt to extend timelines to 2-D graphs or utilize various bump charts \cite{SCHARDONG201899}. This approach, however, requires splitting the sequence into meaningful sub-parts based on content, as demonstrated with categories of user actions \cite{dynatrace2024timelineplot}. Unfortunately, it is not directly possible for the case of log records, since, in order to divide the monitored processes into relevant groups and assign each log record to one of such groups, extensive knowledge about the behavior of the system is required. However, this knowledge is precisely what the proposed method attempts to extract and present.

Our proposal is to integrate a timeline of events with a scatter plot that projects the corresponding semantic content based on similarity, computed directly without the need for embeddings or feature design. The goal is to visualize both temporal and similarity dimensions of the system evolution within a single plot. A direct combination of these techniques is the Time-connected scatter plot \cite{haroz2015connectedscatterplots}, initially used in journalism. A more effective method for encoding time than static labels near data points is the use of colorful segments with discretized time, as introduced by Time-segmented scatter plots \cite{koors2015segmentedscatterplots}. However, connecting points with straight lines can result in high visual complexity. Therefore, in this manuscript, we consider Time Curves \cite{bach2015timecurves} as a more promising approach. Time Curves produce smooth curves where the connection between data points are easier to track and interpret while still allowing for the analysis of the evolution of the system.

\section{Visualizing Logs with Time Curves}
\label{Section_Methodology}

The proposed method is designed to aggregate information from a large collection of system logs into a single and interactive representation. Such representation uses the Time Curves technique to show a series of data points in a 2D space based on the given distance measure. These data points in the plot correspond to system events, which will be spatially closer if their content (\textit{i.e.}, the logs) is similar. This enables both to analyze single log collections and to overlap concurrent system executions to show overall trends and outliers. Furthermore, the representation can be enriched with domain-specific information or summarizations provided by LLMs.

\subsection{Introduction}

In the original \textit{Time Curves} paper \cite{bach2015timecurves}, the authors described several use cases, including analyzes of document changes (such as in Wikipedia article revisions) or video footage. In this paper, the data consists of system logs, where each item is a plain text message, typically including a timestamp along with optional tags and a message payload. While it might appear to be similar to the article revision use case, the sheer cardinality of logs makes it hardly comparable. System logs are often counted in the hundreds of thousands or millions, which makes a projection of such a high number of data points impractical to say the least. Therefore, our proposal must cope with an increase in dimensionality: instead of directly projecting logs as data points for \textit{Time Curves}, it becomes necessary to first aggregate the logs.

\subsection{Workflow overview}

The method (as shown in \cref{Figure_Pipeline}) is composed of (1) data preprocessing, (2) grouping and template extraction, (3) distance computation and (4) projection and summarization. The first stage focuses solely on parsing the input logs (timestamp and other optional tags), joining multi-line logs together and forming log records. The second stage breaks down the preprocessed log records into a sequence of events by grouping consecutive logs and mapping records to templates by performing clustering. The third stage consists of computing distances between templates and aggregating them to get distances between checkpoints (which represent events as sets of templates). The last stage completes the method by projecting checkpoints onto a 2-dimensional plane according to computed distances and joins them in chronological order to produce a \textit{Time Curve} while also enabling annotations and enrichment via LLMs.

\subsection{Preprocessing}

The main difference between a \textit{Time Curve} and a similarity-scatter projection is the fact that the temporal context is kept. Therefore, we extract timestamps from log messages. In addition, we parse other useful information such as severity level tags or particular keywords (\textit{e.g.}, \texttt{Exception}).

While there is no single logging standard, most common logging formats usually include a timestamp as well as a tag representing the severity level before including the payload, \textit{e.g.}, \texttt{"23-09-12 13:01 WARN - Unexpected value ..."}.

We assume that each individual log record includes a timestamp in some form. When such is not found in the line of log file, the message is considered part of the previous record and thus simply appended to it. The parsing of the timestamp is done automatically through regular expression matching and can be easily extended if required.

The severity level is extracted by matching keywords such as \texttt{WARN} or \texttt{INFO} and mapping them to integer constants. The higher the mapping values, the higher the impact they have on the grouping into events. In order to treat the data as a proper time series of log records, we sort records in ascending order of timestamps and therefore do not use the original ordering which might be tampered by batching or bulking mechanisms. 

Once records have been properly parsed, the next step is to group them into events, \textit{i.e.}, split the complete collection of logs into multiple chronologically consecutive sets.

\subsection{Event detection and clustering}
\label{SubsectionGroupingClustering}
Event detection is an inherently difficult problem. Multiple time series analysis or change detection algorithms exist in the literature (\textit{e.g.}, see \cite{ghaderpour2021survey} and \cite{choi2021deep}). Nonetheless, they are typically not designed for discrete $n$-ary valued time series detection, such as in the case of analyzing severity level. For instance, when an error occurs, multiple other errors might follow, or a series of different types of records might be printed. The grouping approach must be able to consider such scenario and group all corresponding logs into a single event.

While prior information about the logging system may be used it to improve the final grouping, the default method takes into account the severity level of log record, the time gaps in between records and the size limits of events.

First, it sets the maximum number of log records for an event, which is estimated based on the desired number of data points (generally between a few tenths and a hundred). Then, it computes the time gaps between consecutive records and determines the positions of the $k$-largest gaps, where $k$ is configurable. If the $(k+1)$-th largest gap is equal to the $k$-th, $k$ is decremented until the gaps are no longer equal or $k$ reaches 0, indicating that more than $k$ largest gaps are equal and time gaps will be ignored.

Next, the method identifies points of severity level change. To ensure robustness, it smooths the extracted severity levels by applying a convolution with a normalized asymmetric window. The decaying and smoothing windows cause the function value to increase rapidly with errors but require several error-free log records to return to the constant level. Such level should vary depending on the system itself but is set by default to the mapping value of \texttt{INFO} plus a small constant. Change points are detected both when the function raises above the threshold (indicating the start of an event) and when it returns below the threshold (indicating the end of an event).

All this information is then combined. The process starts by creating an event with only the first log record. Before adding each next log record, the method checks whether the maximum event size is reached, whether a severity level change point occurs or if the current time gap is within the list of the $k$-largest time gaps. If any of these conditions are met, a new event is created, and log records are added accordingly. This process repeats until the last record is added to an event. The outcome is as follows:
\begin{enumerate}
    \item Errors always trigger a new event unless they are part of a succession of previous errors occurred in the near past, in which case they get bundled together.
    \item Long time intervals with no production of log records result in separate events.
    \item An excessive number of records produced within a small time frame triggers a new event.
\end{enumerate}

After records have been grouped into events, a clustering algorithm \cite{he2017drain} is applied to extract templates. Templates are composed of static parts (such as service names) and variable parts (such as IP addresses). Log records are then mapped to their corresponding templates and the variable parts are masked, thus reducing the cardinality at the expense of exactitude. This is due to (1) sequence order is lost within checkpoints and (2) the masked variable parts may have contained important information. However, without this step, the sheer number of log records would make the projection useless. In general, this results into a coarse-grain aggregation of the major changes that happened within the system. In case a finer-grain analysis is required (\textit{e.g.}, small datasets), the size of the events can be reduced to as small as only one record, thus reducing the amount of information lost due to clustering.
Once clustering has been applied, each event can be represented as a checkpoint, \textit{i.e.}, by a timestamp and a set of all the unique log templates which it contains.

\subsection{Distance properties}
\label{distanceProperties}
A distance function across checkpoints needs to be defined in order to compute a visual projection via Multidimensional Scaling (MDS) \cite{buja2008data}. For such purpose, we want the distance to consider both (1) the similarity between the templates of two checkpoints and (2) the number of templates. While MDS can work with dissimilarity measures that are formally not metrics, additional metric properties of the distance reflect positively in the projection. Our approach is to measure how similar (or distant) two checkpoints are by first defining a similarity measure between log templates and then aggregating it across checkpoints. This is equivalent to finding an appropriate distance for sets of strings, which also accounts for similarity between elements since the context and semantics are of high importance. More formally, we want the function to exhibit the following properties:
\begin{enumerate}
    \item Checkpoints which contain the exact same templates should have a distance of 0. This will result in clusters of equal checkpoints in the final projection.
    \item Checkpoints containing similar templates should have a small distance. Therefore, the distance should account for variations between templates. For instance, using the Jaccard distance \cite{jaccard1901etude} would result in large distance between two sets containing \texttt{"System started rotating logs"} and \texttt{"System finished rotating logs"}, despite their semantic similarities, \textit{i.e.}, both describe the rotation of logs.
    \item Checkpoints which do not contain similar templates should have a large distance, regardless of whether they contain a similar number of templates.
    \item Checkpoints which are subsets of other checkpoints should have a small distance, yet account for the fact that the number of templates varies.
\end{enumerate}
Note that standard set distances like the Jaccard distance are unsuitable since they only take into account whether the elements are exactly the same or not. Moreover, properly weighting the string similarity between templates and the difference in the number of templates between checkpoints is difficult and relates to the fundamental question of quantity versus quality.

\subsection{Distance computation}
\label{SectionDistanceComputation}
We denote checkpoints with lower-case letters and templates with sub-indices, \textit{e.g.}, a checkpoint with $n$ templates is denoted as $x := \{x_0, \dots, x_{n-1}\}$, where all $x_i$ are string log templates. In addition, $|\cdot|$ denotes the cardinality of sets and the length of strings, \textit{e.g.}, $|x| := n$ is the number of templates in checkpoint $x$, $|x_i|$ is the number of characters in the string template.

The similarity between templates is computed via the Levenshtein distance, given its flexibility and available computational optimizations.
However, any distance function designed for variable-length strings can be used as well, \textit{e.g.}, the Longest Common Subsequence or the $q$-gram \cite{ukkonen1985algorithms} distances.

The Levenshtein distance between strings $x_1$ and $x_2$, denoted as $d'(x_1, x_2)$, is defined as the minimum number of single-character edits (insertions, deletions or substitutions) required to change one string into another.

In order to compensate for the fact that the Levenshtein distance  between differently sized strings is governed by the length difference rather than the actual string contents, we normalize by the maximum length between strings. And to keep the distance between $[0,1]$, we also normalize by the weights used in the Levenshtein function: \\
$$d(x_1,x_2) = \frac{d'(x_1,x_2)}{\max(w_{ins}, w_{del}, w_{sub}) \cdot \max(|x_1|, |x_2|)}.$$

We include the normalization by weights to further generalize the proposed distance, allowing a custom weighting to favor certain types of editions if prior information about the logging system exists. However, since weights are one by default, no weight normalization is required.

The next step is aggregation. Our proposal consists of iterating through all templates of one checkpoint to find the closest template in the other checkpoint and collect the distances.

The arithmetic mean of all the collected distances between templates of checkpoint $x$ and $y$ is used to stay within the range $[0,1]$:
 \begin{equation}
     D'(x,y) := \frac{1}{|x|} \sum_{i=0}^{|x|-1} \min_{j = 0, 1, \dots,|y|-1} d(x_i, y_j).
     \label{eq: Directed Semimetric}
 \end{equation}

For $D'(x,y)$ we iterate and average over templates of $x$ whereas for $D'(y,x)$ -- over $y$, therefore $D'$ is not symmetric. However, solving with $D(x,y) := D(y,x) := \frac{D'(x,y) + D'(y,x)}{2}$ disregards cardinalities of checkpoints, making the measure too sensitive to checkpoints with few templates. In order to make the distance more robust to the number of templates, the following logarithmic weighting was chosen:
\begin{equation}\label{eq: D(x,y) definition}
    D(x,y) := \frac{\log_2 |x| \cdot D'(x,y) + \log_2 |y| \cdot D'(y,x)}{\log_2 (|x|\cdot |y|)}
\end{equation}

The proposed distance function is a semimetric (and not a metric) since the triangular inequality is violated, however, all other properties hold \cite{supplementaryMaterial}. Note that this semimetric requires significant computation since it comprises multiple quadratic terms, so actual calculation differs from definition. Further details on computational optimizations are available in \cite{supplementaryMaterial}.

\subsection{Changes to the original \textit{Time Curves}}

We included multiple extensions to the \textit{Time Curves} approach in order to better suit the use case of analyzing complex logs of software systems.

From the user interface, we replaced the color scale in order to account for time. In addition, we included an animation option to draw the \textit{Time Curve} segment-by-segment from start to end. This helps to understand the evolution of the system by slowly adding checkpoints, especially as projections get more complex. Example animations can be found in \cite{supplementaryMaterial}. Lastly, we included the coefficient of determination ($R^2$) as an indicator of the projection quality.

From the data augmentation perspective, the data points are enriched by including the log template messages contained within as annotations that can be opened and closed via clicking. This enables the user to explore and compare the log messages of different instants of the system. Furthermore, this facilitates further analysis via \textit{e.g.}, LLMs.

From the functionality perspective, we included the possibility of overlaying multiple curves together, which can be used to compare multiple similar systems at once (\textit{e.g.}, \cite{nagaraj2012logsdiffing}). More details can be found in \cref{SubsectionDiffing}.

\section{Evaluation}
\label{Section_ResultsAndDiscussion}

\begin{figure*}[!htbp]
   \includegraphics[width=\textwidth]{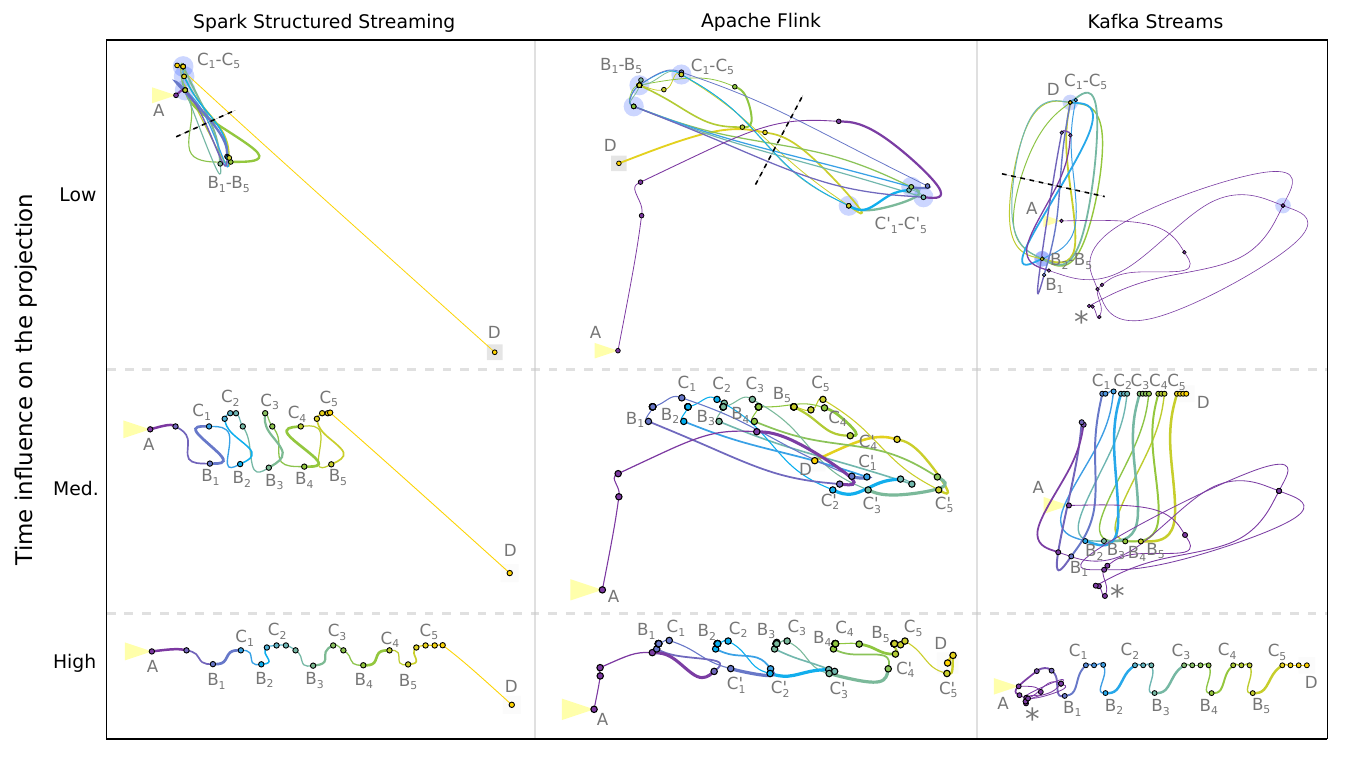}
\caption{Projections of the collected logs for three stream processing framework executions and their unfolding according to the ratio of similarity and time influences. The annotated labels (\texttt{A} to \texttt{D}) represent checkpoints of interest for the analysis of the system, namely (\texttt{A}) startup, (\texttt{B1-B5}) failure injections, (\texttt{C1-C5}) recovery and (\texttt{D}) shutdown. In addition, \texttt{C1'-C5'} mark normal processing after recovery and a \texttt{\large$\ast$} represents a singular checkpoint in Kafka Streams. }
\label{Figure_ThreeFrameworks}
\end{figure*}

In order to evaluate the method, we conducted multiple experiments, focusing on explainability and scalability. First, to confirm the expected behavior of the proposed approach, an experiment was repeated under the same conditions on three different stream processing frameworks, and the output logs were analyzed. The objective is to achieve a consistent analysis across all frameworks despite their logging differences.
We also analyzed a publicly-available collection of logs first manually and then leveraged the summarizing capabilities of LLMs for single and pairwise checkpoints. This is aimed at showing how the visual interpretation of the \textit{Time Curves} can be used to gather insights while also enriching with context information. In addition, we analyzed a proprietary log dataset of three concurrent systems to showcase the multi-curve visualization.
Lastly, for the case of scalability, we measured the performance of the method for a variety of publicly-available log datasets of different sizes.

\subsection{Stream processing analysis}
\label{Section_StreamProcessingAnalysis}

While many public log datasets are available in the literature, they are hardly annotated or explained, thus making it very difficult to analyze without bias. Therefore, we reused the log data of our previous research on fault recovery \cite{vogel2024comprehensive}, which includes three popular open-source stream processing frameworks, namely Apache Flink \cite{Carbone2015}, Spark Structured Streaming \cite{Armbrust2018} and Kafka Streams \cite{Wang2021}. This provides an explainable collection of logs from different applications executing the same workload, enabling a comparison of the logs. The expectation is that the proposed method results in a similar analysis outcome for all cases.

The logs include a 72 minute execution of the ShuffleBench benchmark \cite{henning2024shufflebench} where 5 failures were automatically injected every 12 minutes, starting at the 12-minute mark. Default parameters of the event detection algorithm were used, except for the case of Kafka Streams, which required to fine-tune the time granularity due to the fact that the collector node received nearly no logs from the failing nodes. Further details on the experimental setup are available in \cite{vogel2024comprehensive, supplementaryMaterial}.

\Cref{Figure_ThreeFrameworks} shows the \textit{Time Curves} projections of each of the executions of the frameworks in three different ways: with a low, medium and high importance of the time dimension. This is provided because of the limitations of showing interactive plots within a static manuscript. Animations can be found in \cite{supplementaryMaterial}. In addition, \cref{Figure_ThreeFrameworks} also includes the labeling of the processing stages that take place throughout the execution, namely (\texttt{A}) startup, (\texttt{B1-B5}) failure injections, (\texttt{C1-C5}) fault recovery and (\texttt{D}) shutdown of the system.

Besides the starting and shutdown sequences, notice how the upper row of \cref{Figure_ThreeFrameworks} (which corresponds to the projection based exclusively on similarity) shows a cyclic behavior. All three frameworks exhibit a linearly separable structure that splits failures from recoveries or normal processing, which is shown with a manually-added dashed-black line. This becomes more apparent as the curve gets unfolded by time in the middle and lower rows. Due to the proximity between the labels of failure and recovery in Flink, the normal processing checkpoints are also highlighted through labels (\texttt{C1'-C5'}).

Example logs for each of the checkpoints can be found in \cite{supplementaryMaterial}. Both the startup (\texttt{A}) and shutdown (\texttt{D}) can be easily identified since they contain unique messages. This is the case for all frameworks except for Kafka Streams, where there is no actual record corresponding to the shutdown sequence. In fact, \texttt{D} contains only generic information messages regarding normal processing, which is no different from the previous checkpoints that take place after recovery in \texttt{C5}. This is due to how the execution was conducted and the architecture of Kafka Streams. In particular, the lack of a manager node and how the nodes are terminated after the benchmark execution result in no specific logs for the shutdown.

Similarly, both Spark and Flink curves transition much faster from startup to the first injected failure (\texttt{A} to \texttt{B1}), with Spark including a single checkpoint in between and Flink four additional ones. Note that the word checkpoint is used as a data point in the figure, rather than the checkpointing process in stream processing frameworks. On the contrary, Kafka Streams is separated into two different patterns: an initial loop right after startup and the main loop afterwards. The first one is due to the initial partition assignments and balancing happening before stable processing is achieved. The second one is due to the recurrent failure injections. Interestingly, it can be seen that the projection places the cluster of failures (labels \texttt{B1-B5} close to the tightly coupled checkpoints of the initial partition assignment (marked with\ \texttt{\large$\ast$}). This is due to the large overlap in terms of logs between the initial partition assignment and the failure injections, since after a failure a re-balancing and new partition assignment is triggered \cite{vogel2024comprehensive}.

After the first injected failure and recovery (\texttt{B1} and \texttt{C1}), all projections exhibit a cyclic behavior. This can be clearly seen in the upper row, where failures and recoveries are clustered separately. Due to the number of different logs and templates, small variations occur between each of the failures and the recoveries, making the checkpoints not completely aligned.

A higher influence of time (\textit{i.e.}, bottom row) provides a more linear representation of the sequence of events. In fact, the periodicity of the system (every 12 minutes) can be seen in terms of constant distance between the failures. The time to recovery can also be inferred, showing that in most cases it also remains quite constant, although different for each framework. Notice that further analysis is possible by increasing or decreasing the checkpoint granularity. 

These results confirm our expectations based on our previous research \cite{vogel2024comprehensive}, particularly by capturing the cyclic structure of failures and recoveries. It also provides additional information such as the unexpected startup loop of the Kafka Streams execution.

\subsection{Public dataset analysis}
\label{Section_PublicDatasetAnalysis}

In this section, we analyze a publicly-available log dataset and compare it with manual inspection. The dataset in particular is the Zookeeper log collection from the LogHub repository, which contains roughly 75,000 records. Apache Zookeeper is a framework that enables distributed process coordination while providing mechanisms for fault tolerance and state preservation. The collection of logs contains records pertaining to the communication of nodes, leader election procedures, client requests, \textit{etc}. In this particular setting, three machines act as servers whereas the rest are clients.

\begin{figure}[tbp]
   \includegraphics[scale=0.6]{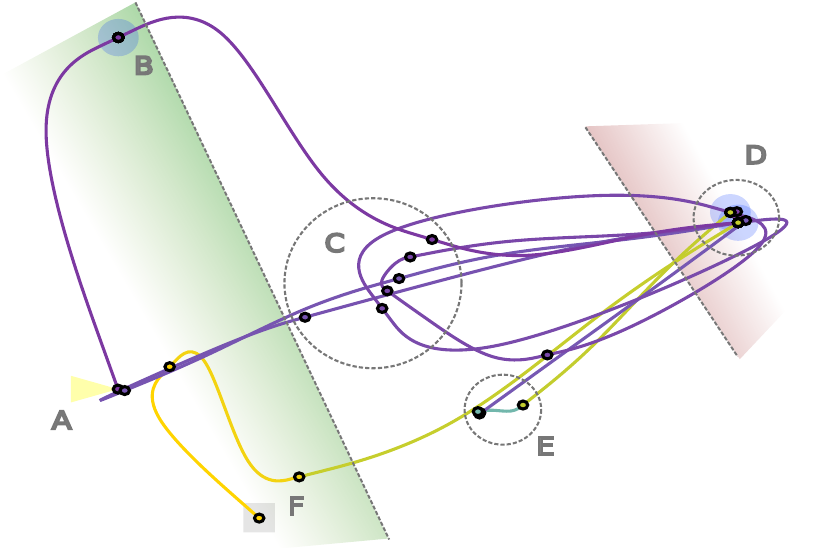}
\caption{Identified sections of interest corresponding to the Zookeeper log dataset. Labels \texttt{A} to \texttt{F} represent points of interest. Two planes marked with a gradient color separate the projection into left, middle and right sections, corresponding to different stability stages of the system.}
\label{FigZookeeperAnalysis}
\end{figure}

\Cref{FigZookeeperAnalysis} shows the \textit{Time Curve} for the dataset along with identified sections of interest, including labels \texttt{A} to \texttt{F} and the left, middle and right plane sections. These will be used to describe the processes taking place throughout the logs.

In general, the Zookeeper service begins with startup, where multiple rounds of elections take place to choose the leader server. These produce logs about communication between the three existing servers. Once a leader is elected, clients may start connecting with or without success. In addition, servers may crash due to unknown circumstances and therefore trigger new elections. This process continued for roughly 26 days according to the collected logs.

The analysis can be summarized by splitting the projection into the following three planes:

\begin{enumerate}
    \item \textbf{Left section} includes startup (\texttt{A}), initial configuration attempts (\texttt{B}) and end (\texttt{F}), but contains no client requests and depicts that the system is not in a ready state.
    \item \textbf{Middle section} shows relative stability stages in which the system is servicing client requests, however it is intertwined with election procedures either routinely (\texttt{C}) or due to occasional crashing of a server (\texttt{E}).
    \item \textbf{Right section} depicts high stability stages in which no election procedures are required, clients are connecting and requests are being fulfilled (\texttt{D}). No server crashes take place during these checkpoints.
\end{enumerate}

In addition, a cyclic pattern between the relative and high stability stages (\texttt{C} and \texttt{D}) can be observed, both in terms of events and time (see drastic change in color between data points in \texttt{D}, which denotes a large time gap).

Further inspection of the individual labels may be possible:

\begin{enumerate}
    \item Label \texttt{A} corresponds to the startup of the Zookeeper service. This involves log records such as the main thread starting the application and reading configuration files. Note that \texttt{F} (and its surrounding data points) are close to \texttt{A} due to lack of client connections. In addition, the error \texttt{"Cannot open channel to ... at election address"} prevents the election to complete successfully, thus triggering further elections processes and therefore resembling the startup sequence. Such error gets resolved and the election completes at \texttt{F}.
    \item Label \texttt{B} shows an interesting case: after startup and initial election, the system enters a repetitive phase where servers are trying to communicate with each other but are unsuccessful. An unspecified error results in broken connections for all servers, \textit{i.e.}, \texttt{"Connection broken for id ... error = "}. This behavior takes place for roughly half an hour and is responsible for almost 75\% of all the log records in the dataset. Note that despite the overrepresentation of \texttt{B}, the proposed method is still able to capture several other behaviors.
    \item The cluster of data points \texttt{C} contains the first data point where client connections begin to happen. Occasional elections take place as well. The variability between the data points is caused by not all checkpoints containing the same reasons for server crashes, as well as additional records such as in regards to the creation of log files.
    \item Label \texttt{D} shows the stage of the system where highest stability is achieved without server crashes or election procedures. Only client connections take place. Again, the variability is explained by small differences such as occasional exceptions that do not result in crashes.
    \item Label \texttt{E} shows a large time gap up until \texttt{D} of nearly 8 days of limited activity. The connection of clients resumes with occasional crashing prior to returning back to the stable stage in \texttt{D}. The main difference between \texttt{E} and the cluster of data points at \texttt{C} is that new errors appear regarding the Zookeeper service not running temporarily.
\end{enumerate}

\subsection{Enrinchment via LLM}
\label{Section_EnrichmentViaLLM}

When the collection of logs comprises a large number of templates (\textit{e.g.}, hundreds of templates in a single checkpoint), manual inspection as in \cref{Section_PublicDatasetAnalysis} becomes difficult due to the amount of information contained. As each point is represented by string templates, given context and domain source, LLMs can be used to further summarize the content of a single checkpoint as well as compare different data points of the \textit{Time Curve}. For the following analysis, we used Microsoft's Copilot model which is based on GPT-4 \cite{narayanaswamy2024evolution}.

\subsubsection{Single point summarization}

The objective is to explain the templates contained in a single checkpoint. This can be helpful for a variety of reasons, \textit{e.g.}, adding context to the meaning of log records, especially when users are not familiar with the semantics of the logging format, further summarizing a large number of templates into few sentences or reasoning about particular log messages, \textit{etc}. 

Two arbitrary checkpoints are summarized from stages \texttt{C} and \texttt{D}. The model explained stage \texttt{C} as ``\textit{The logs indicate that the ZooKeeper server is actively managing client connections and sessions, handling leader elections, initializing and configuring the server, and synchronizing data with followers. The server also deals with various warnings and errors, such as unexpected shutdowns, broken connections, and session expirations}''. Note that this captures our previous manual explanations. When further inquired about the gravity of the errors, the model indicates that they should be investigated due to their severity, and that they are not part of what can be considered normal processing.
    
On the contrary, in \texttt{D} (stability stage), the model explained that ``\textit{The logs indicate that the ZooKeeper server is actively managing client connections and sessions. It handles new connection attempts, establishes and terminates sessions, and processes session expirations. The server also deals with connection issues, such as end-of-stream exceptions and connection requests from outdated clients.}''. This is in line with our analysis on the previous section. Note that the end-of-stream exceptions do not result in server crashes. Further investigation about the seriousness of the end-of-stream exceptions result in the model explaining that they are relatively common and not problematic, as opposed to the checkpoint in \texttt{C}.

\subsubsection{Pairwise point comparison}

The objective of pairwise comparison is to describe the differences (or similarities) between two given checkpoints. This is helpful when comparing the behavior of the system over time or to correlate visual differences in the \textit{Time Curve} with the actual set of logs. 

First, we selected two close checkpoints from \texttt{D} (stable processing) to find out if the difference is significant. The model explains that both checkpoints are nearly identical except for the presence of a single additional log record related to a \texttt{KeeperException} where an attempt to create an already existing node was made. When further inquiring, the model describes the exception as a non-critical error which could be handled more gracefully. This is in agreement with our manual investigation, where we found out that no server crashes occur during the high-stability stage (Label \texttt{D}).
    
Second, we selected \texttt{C} and \texttt{E} from the unstable procession stages. On one hand, regarding the similarities, the model explains that both are ``\textit{... capturing various operational events and issues related to leader election, session management, and client connections}''. On the contrary, regarding the differences, first the model returns a generic (yet true) summarization, namely ``\textit{the first set of logs includes more detailed information about the server environment and configuration, while the second set includes more details about the election process and issues encountered during synchronization and session management.}'' along with a list of the differences depending on their severity level (\textit{i.e.}, variations by \texttt{INFO}, \texttt{WARN}, \textit{etc.}). When a more specific explanation for the set of \texttt{WARN} logs is requested , the model replies ``\textit{These logs highlight issues related to leader election and synchronization, such as unexpected zxid values and exceptions when following the leader, as well as a specific session closure due to the server not running}'', which is in line with our manual analysis where we found out that the Zookeeper service is not running temporarily.

These examples show that the information conveyed within checkpoints in the shape of template representations seems sufficient to produce valuable insights which are in line with potentially time-consuming manual inspections.

\subsection{Multiple curve analysis}
\label{SubsectionDiffing}

The logging output of software systems can widely differ. However, some instances of the same application may be running concurrently, thus increasing the likelihood of their output being similarly structured. In this section, we use our approach to overlay the projection curves of multiple application instances to identify outliers or branching differences.
\begin{figure}[h]
    \centering
    \includegraphics[scale=0.45]{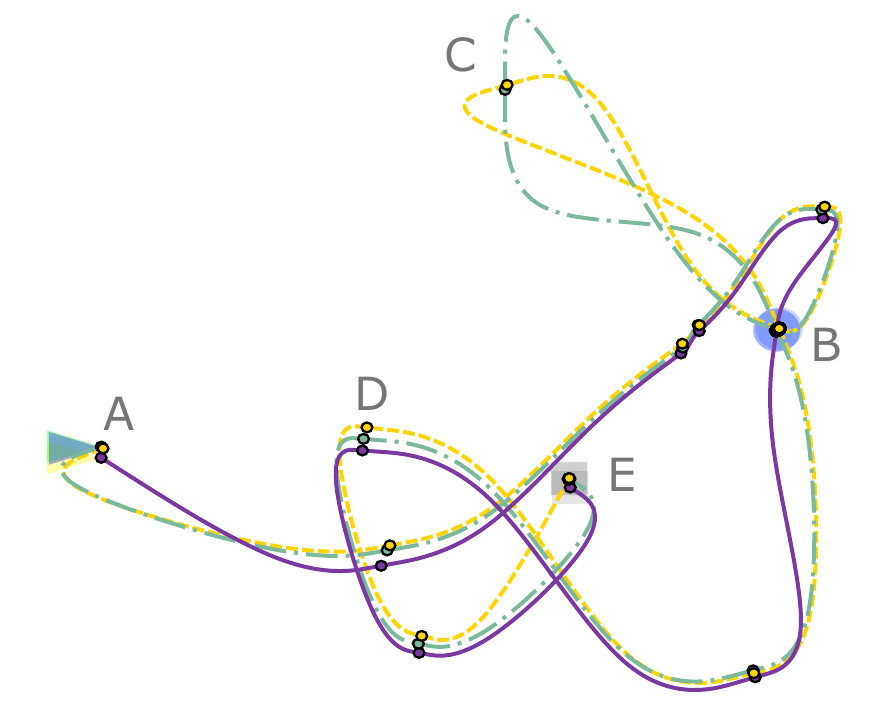}
\caption{Multiple curve analysis of three instances of the same application. Each system is represented with a different color and line style to ease visualization. In addition, due to the fact that different colors are used to represent different systems, the time dimension is no longer represented besides the order of checkpoints. Labels \texttt{A} to \texttt{E} represent stages of interest.}
\label{diffingCurves}
\end{figure}

To illustrate a multiple curve analysis, we gathered log collections corresponding to three instances of a simple server application that receives connections from clients. The application was monitored from launch to shutdown, with the main processing stage being listening to remote connections. \Cref{diffingCurves} shows the \textit{Time Curves} of the three log collections overlaid on top of each other. Labels \texttt{A} to \texttt{E} represent the following stages:

\begin{itemize}
    \item Label \texttt{A} corresponds to the startup stage. It can be observed that all three curves begin at nearly the exact location, due to the initialization being identical except the small variations, caused by \textit{e.g.}, different execution paths or process IDs. These cause a very small distance that enables to differentiate between exactly equal executions and semantically equal executions.
    \item The startup continues until \texttt{B} where the three systems enter the processing stage. The applications are listening to remote connections. Notice that each one receives a variable number of connections yet they are visually clustered in \texttt{B}.
    \item Until stage \texttt{C}, processing had remained equal for all applications. However, it can be observed that two of the curves, namely the yellow dashed and the dot-and-dashed green, enter an outlier checkpoint in \texttt{C} before returning to the normal processing (Label \texttt{B}). In fact, by examining the logs corresponding to \texttt{C} it can be seen that the two curves threw a \texttt{"An illegal reflective access operation has occurred"} followed by error logs, which do not occur in the solid purple curve. 
    \item After normal processing has resumed, all applications begin the shutdown stage from \texttt{D} to \texttt{E}.
\end{itemize}

The analysis of \cref{diffingCurves} shows at first glance outlier behavior from two out of the three executions without requiring to manually compare the log collections. In a similar fashion, hundreds of curves can be overlaid at once, resulting in a coarser-grain representation, with common processing paths becoming thicker as a result of the bundling of segments, as well as the outliers being identifiable as unique lines.

\subsection{Scalability of the approach}


As Dynatrace deployments keep growing, the efficiency of log analytics becomes extremely important, especially given the scale of the log ingest. Therefore, in this section, we show how the method scales both in terms of size and complexity of the input data. By size we refer to number of input log lines (which approximately results in a similar number of log records), whereas for complexity we mean the distribution of the lengths of the records and the number of extracted templates. On one hand, longer log lines have an impact on performance due to the intrinsic quadratic term of the Levenshtein distance. On the other hand, more templates also result in a performance penalty because of the pairwise template comparison between checkpoints. While this can be become a potential bottleneck, further optimizations may be applied to improve performance as a trade off for accuracy, such as using a computationally cheaper string metric as replacement of the Levenshtein distance (\textit{e.g.}, the q-gram distance \cite{ukkonen1992approximate}) or limiting the number of extracted templates during the clustering stage. 

All experiments described in this section were averaged over 10 runs. Further setup details can be found in \cite{supplementaryMaterial}.

\begin{figure}[htbp]
   \includegraphics[scale=0.395]{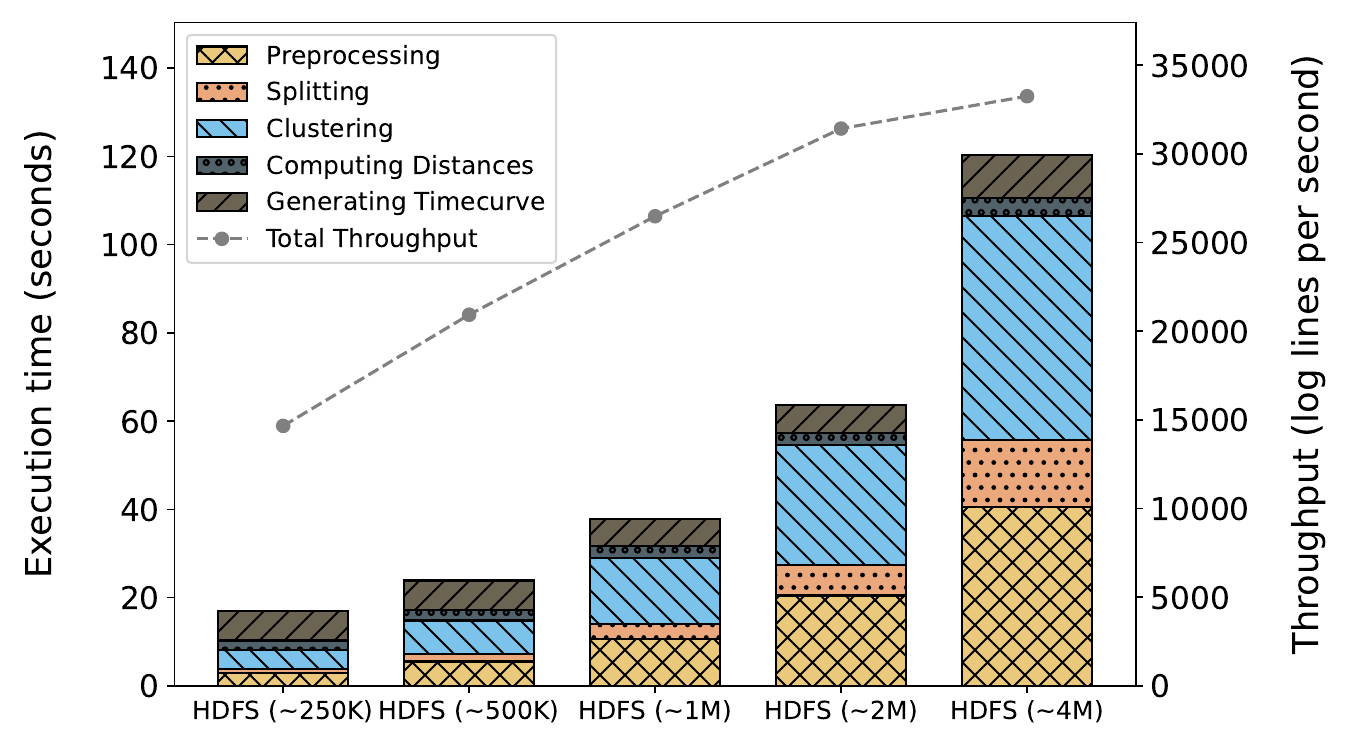}
\caption{Average runtime and throughput as a function of the number of log messages of the Hadoop file system logs. Note that x-axis is in $\log_2$ scale. The left y-axis represents average runtime in seconds using stacked bars, whereas the second y-axis represents average throughput using a dashed line. The proportion of time required by each of the stages is represented as part of the stacked bars.}
\label{figScalabilityHDFS}
\end{figure}

\Cref{figScalabilityHDFS} shows the average runtime and throughput in terms of processed records per second for an increasing number of logs. Note that the dataset source is fixed to maintain the distribution of the log records. As can be observed, peak performance is achieved with the largest dataset containing four million log records at a processing rate of roughly 35,000 log records per second (with a single core). The smaller datasets do not contain enough log lines to pay off for (1) constant initialization factors, such as importing of libraries, (2) multidimensional projection, which depends on number of checkpoints and goodness of fit rather than number of log records, (3) curve annotation, \textit{etc.} These penalties get smoothed out as the actual core processing (grouping, template extraction and distance computation) takes over when the number of records increases. 
It can be concluded that the runtime (and throughput) of the method is linear at worst given that the x-axis is in logarithm scale.

\begin{figure}[htbp]
   \includegraphics[scale=0.395]{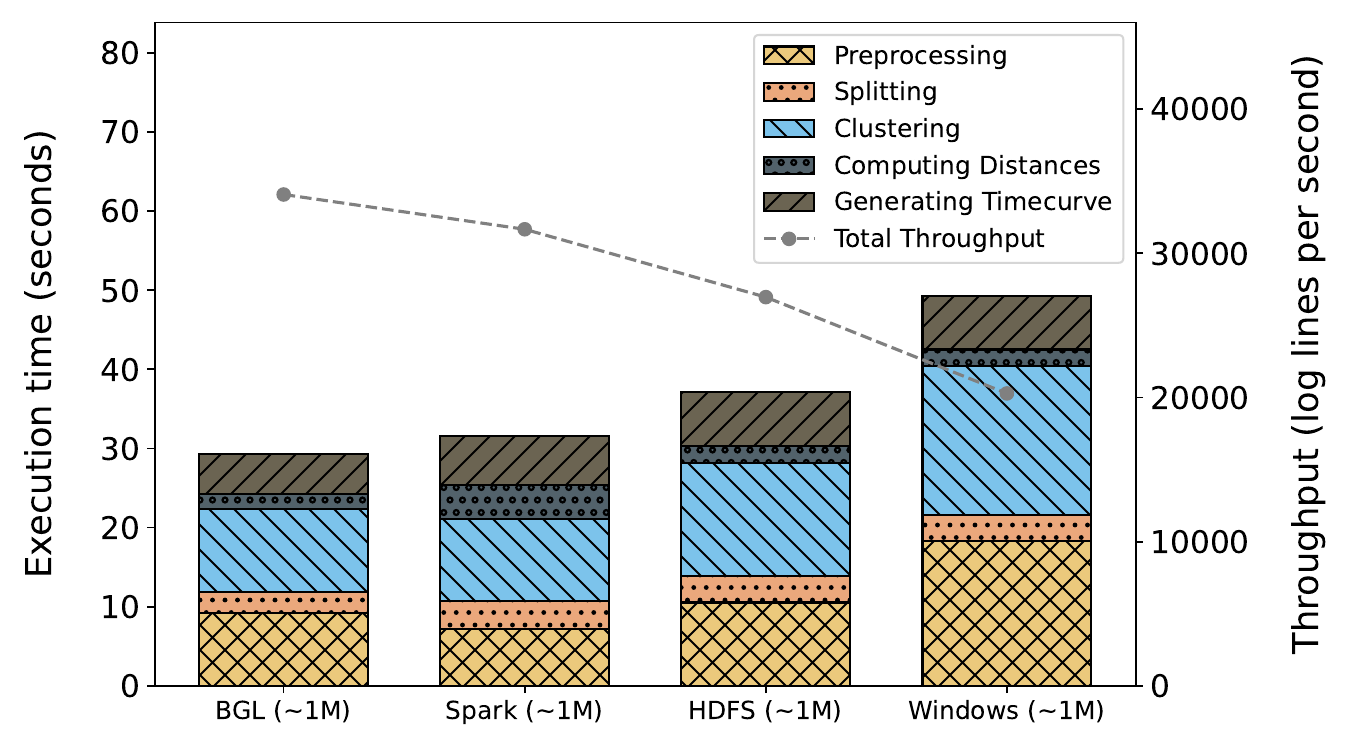}
\caption{Average runtime and throughput as a function of the dataset. The left y-axis represents average runtime in seconds using stacked bars, whereas the second y-axis represents average throughput using a dashed line. The proportion of time required by each of the stages is represented as part of the stacked bars.}
\label{figCoverageDatasets}
\end{figure}

\Cref{figCoverageDatasets} shows the average runtime and throughput for one million log lines of different datasets from the LogHub repository \cite{zhu2023loghub}, including Spark, HDFS, BGL and Windows. The number of log lines was fixed to focus only on the variability of the dataset. Runtime proportions remain relatively stable throughout the datasets, however not constant, which can be explained by the differences in log line lengths (average varying between 102 and 140) and the number of extracted templates (from 35 to 250), the distribution of similar records, \textit{etc}. It should be noted that when the number of templates is extremely large (\textit{e.g.}, above 500), the throughput can decrease significantly. This was observed with the Thunderbird dataset (also from the LogHub collection). The equivalent one million log lines resulted in roughly 1,600 templates and the throughput dropped by a factor of two. Nonetheless, these extreme cases can be handled by fine-tuning the clustering algorithm to achieve fewer templates and by limiting the maximum length of the templates used for the distance computation.

\section{Conclusions and future work}
\label{Section_Conclusions}

In this manuscript, we have shown that our proposal is able to summarize and represent large collections of log messages while scaling efficiently and providing meaningful insights. This is achieved via the following contributions:
\begin{enumerate}
    \item An automated grouping methodology for detecting events in a collection of logs based on cumulative information (such as timestamps and severity level) without requiring any prior knowledge.
    \item The incorporation of clustering step to reduce dimensionality and enable a meaningful representation, while also improving the runtime complexity of the method.
    \item A semimetric function that can measure distance between sets of log templates while considering both the semantics and the cardinality of the templates.
    \item Several extensions to the \textit{Time Curves} visualization technique, including multiple curve comparison, enriching data points with logging information or animations.
    \item The experimental approach, serving as a reference for validating analysis of log datasets, which can be reused and compared with future methods.
\end{enumerate}

Furthermore, our results show that the approach is in fact able to explain the logs collected from three different stream processing frameworks and how the similarity projection reveals further properties of the framework logging which had not been foreseen (such as the resemblance between startup and failure recovery in Kafka Streams in \cref{Section_StreamProcessingAnalysis}). In the same line, we have also analyzed a public dataset and then compared our manual explanations with the automated results from the proposed method.

From the computational perspective, we have shown that the proposed method can scale efficiently in terms of number of input records as well as complexity of the datasets, thus making the method suitable for log collections of up to hundreds of thousands of log records with low-latency of few seconds. While more fine-grained analysis can be achieved with more computation, low-latency is typically a requirement within the industry.

It should be noted that in this paper, we have focused on analyzing complete log datasets. However, working with subsets according to some criteria (splicing datasets by a desired property) can theoretically produce more focused and tailored insights, \textit{e.g.}  analyzing only a certain component of an application such as the \texttt{ExecutionGraph} in Apache Flink.

In conclusion, we believe the proposed approach offers a holistic method that can ease up system log analysis, thus speeding up analysis of system-wide problems and performance bottlenecks, identification of security threats, \textit{etc}.

In further research, we plan to address the following:

\begin{enumerate}
    \item Further explore the use of LLMs by utilizing both alignment techniques (\textit{e.g.}, prompting) and more complex methods such as Retrieval-Augmented Generation \cite{lewis2020retrieval} while leveraging multimodal models or fine-tuning them.
    \item Speed up processing of even larger datasets by exploiting the parallelism potential,  such as in the preprocessing, event detection or clustering.
    \item The event detection mechanism may benefit from exploring the use of (1) Fast Fourier Transform for faster convolutions, (2) adaptive thresholds to refine the boundary between begin and end of events and (3) additional information such as performance metrics.
\end{enumerate}

\bibliography{bibliography} 

\end{document}